# THE DEPENDENCE ON THE INITIAL STATES AND THE TRANSITIVITY OF THE REGULAR AUTONOMOUS ASYNCHRONOUS SYSTEMS

Şerban E. VLAD

Oradea City Hall, serban_e_vlad@yahoo.com

***Abstract.*** *The asynchronous systems are non-deterministic real time, binary valued models of the asynchronous circuits from electronics. Autonomy means that there is no input and regularity means analogies with the (real) dynamical systems. We introduce the concepts of dependence on the initial states and of transitivity for these systems.*

## 1. INTRODUCTION

The asynchronous systems model non-deterministically the asynchronous circuits from electronics. Under the form from this paper, they continue the switching theory from the 50's and the 60's. The autonomous systems are systems without input and the regular systems are systems obtained by analogy with the dynamical systems. Our aim is to define the dependence on the initial states and the transitivity of the regular autonomous asynchronous systems. We give also a brief introduction in this theory.

The $R \to \{0,1\}$ functions (real time, binary values) are not studied in literature and our work is original. The future interests are related with continuing the analogies with the dynamical systems.

## 2. PRELIMINARIES

*Notation* 2.1 The set $B = \{0,1\}$ is the binary Boole algebra. Its usual topological structures are the discrete topology and the rough topology.

*Definition* 2.2 The sequence $\alpha : N \to B^n, \alpha(k) \overset{not}{=} \alpha^k$, $k \in N$ is called progressive if the sets
$$\{k \mid k \in N, \alpha_i^k = 1\}$$
are all infinite, $\forall i \in \{1,...,n\}$. We denote the set of the progressive sequences by $\Pi_n$.

*Definition* 2.3 For the function $\Phi : B^n \to B^n$ and $\nu \in B^n$ we define $\Phi^\nu : B^n \to B^n$ by $\forall \mu \in B^n$,
$$\Phi_i^\nu(\mu) = \overline{\nu_i}\mu_i \oplus \nu_i \Phi_i(\mu), i = \overline{1,n}.$$

*Definition* 2.4 The functions $\Phi^{\alpha^0 ... \alpha^k} : B^n \to B^n$, $k \in N$ are iteratively defined for $\alpha \in \Pi_n$ by
$$\Phi^{\alpha^0 ... \alpha^k \alpha^{k+1}}(\mu) = \Phi^{\alpha^{k+1}}(\Phi^{\alpha^0 ... \alpha^k}(\mu)), k \in N.$$

*Remark* 2.5 $\Phi^{\alpha^0 ... \alpha^k}$ shows how the asynchronous discrete time iterations of $\Phi$ are made: if $\alpha_i^j = 1$ then $\Phi_i(\Phi^{\alpha^0 ... \alpha^{j-1}}(\mu))$ is computed and if $\alpha_i^j = 0$, then $(\Phi^{\alpha^0 ... \alpha^{j-1}})_i(\mu)$ is computed. In the special case that $\alpha^k = (1,...,1), k \in N$, $\Phi^{\alpha^0 ... \alpha^k}$ are the iterations of a discrete time Boolean dynamical system.

*Notation* 2.6 We denote by $\chi_A : R \to B$ the characteristic function of the set $A \subset R$.

*Notation* 2.7 We denote by $Seq$ the set of the real sequences $t_0 < t_1 < t_2 < ...$ unbounded from above.

*Definition* 2.8 The functions $\rho : R \to B^n, \forall t \in R$,
$$\rho(t) = \alpha^0 \chi_{\{t_0\}}(t) \oplus ... \oplus \alpha^k \chi_{\{t_k\}}(t) \oplus ...$$
$\alpha \in \Pi_n, (t_k) \in Seq$ are called progressive and their set is denoted by $P_n$.

*Definition* 2.9 The function

Şerban E. VLAD

$$\Phi^\rho(\mu,t) = \mu\chi_{(-\infty,t_0)}(t) \oplus \Phi^{\alpha^0}(\mu)\chi_{[t_0,t_1)}(t) \oplus ...$$
$$... \oplus \Phi^{\alpha^0...\alpha^k}(\mu)\chi_{[t_k,t_{k+1})}(t) \oplus ...$$

is called flow, motion, or orbit (of $\mu \in \boldsymbol{B}^n$).

*Definition* 2.10 The set
$$Or_\rho(\mu) = \{\Phi^\rho(\mu,t) | t \in R\}$$
is also called orbit (of $\mu \in \boldsymbol{B}^n$).

*Definition* 2.11 The universal regular autonomous asynchronous system that is generated by $\Phi$ is:
$$\Xi_\Phi = \{\Phi^\rho(\mu,\cdot) | \mu \in \boldsymbol{B}^n, \rho \in P_n\};$$
$\mu \in \boldsymbol{B}^n$, $x \in \Xi_\Phi$ are called initial state and state.

*Remark* 2.12 The asynchronous systems are, in general, non-deterministic due to the fact that the modeled circuits have unknown (maybe variable) parameters, such as initial states, delays, temperature, the tension of the mains etc. Non-determinism is produced, in the case of $\Xi_\Phi$, by the fact that the initial state $\mu$ and the way $\rho$ of iterating $\Phi$ are not known, giving the so-called (non-initialized) unbounded delay model of computation of the Boolean function $\Phi$. Universality means the greatest in the sense of the inclusion.

*Theorem* 2.13 [5] The set $\Omega_n$ of the bijections $H: \boldsymbol{B}^n \to \boldsymbol{B}^n$ satisfying
i) $H(0,...,0) = (0,...,0)$,
ii) $H(1,...,1) = (1,...,1)$,
iii) $\forall k \geq 2, \forall \mu^1 \in \boldsymbol{B}^n,..., \forall \mu^k \in \boldsymbol{B}^n$,
$\mu^1 \cup ... \cup \mu^k = (1,...,1) \Leftrightarrow H(\mu^1) \cup ... \cup H(\mu^k)$
$= (1,...,1)$
is a group relative to the composition of the functions.

*Definition* 2.14 Let be $\Phi, \Upsilon: \boldsymbol{B}^n \to \boldsymbol{B}^n$ and the bijections $H: \boldsymbol{B}^n \to \boldsymbol{B}^n$, $H' \in \Omega_n$. If
$$\forall \nu \in \boldsymbol{B}^n, \ H \circ \Phi^\nu = \Upsilon^{H'(\nu)} \circ H,$$
then $\Xi_\Phi$ and $\Xi_\Upsilon$ are called conjugated (or equivalent).

*Remark* 2.15 We use in this paper state portraits (also called phase portraits or state transition graphs). For example in Fig. 1 we have the function $\Phi: \boldsymbol{B}^2 \to \boldsymbol{B}^2$, $\forall \mu \in \boldsymbol{B}^2, \Phi(\mu_1,\mu_2) = (\overline{\mu_2} \cup \mu_1\mu_2, \overline{\mu_1} \cup \mu_1\mu_2)$ and in Fig. 2 we have the constant function $\Phi: \boldsymbol{B}^2 \to \boldsymbol{B}^2$, $\forall \mu \in \boldsymbol{B}^2, \Phi(\mu_1,\mu_2) = (1,1)$. When a coordinate $\mu_i$ is underlined, this means that $\Phi_i(\mu) \neq \mu_i$ and when a coordinate $\mu_i$ is not underlined, this means that $\Phi_i(\mu) = \mu_i, \forall i \in \{1,...,n\}$. The arrows show the increase of time.

## 3. MAIN DEFINITIONS AND RESULTS

*Definition* 3.1. The system $\Xi_\Phi$ is said to be p-independent, respectively n-independent on the initial states if $\exists \mu \in \boldsymbol{B}^n, \forall \mu' \in \boldsymbol{B}^n, \exists \rho \in P_n, \forall t \in \boldsymbol{R}$,
$$\bigcup_{\xi \in (-\infty,t]} \rho(\xi) = (1,...,1) \Rightarrow \Phi^\rho(\mu,t) = \Phi^\rho(\mu',t),$$
$\exists \mu \in \boldsymbol{B}^n, \forall \mu' \in \boldsymbol{B}^n, \forall \rho \in P_n, \forall t \in \boldsymbol{R}$,
$$\bigcup_{\xi \in (-\infty,t]} \rho(\xi) = (1,...,1) \Rightarrow \Phi^\rho(\mu,t) = \Phi^\rho(\mu',t).$$

$\Xi_\Phi$ is by definition p-dependent, respectively n-dependent on the initial states if
$\forall \mu \in \boldsymbol{B}^n, \exists \mu' \in \boldsymbol{B}^n, \exists \rho \in P_n, \exists t \in \boldsymbol{R}$,
$$\bigcup_{\xi \in (-\infty,t]} \rho(\xi) = (1,...,1) \ \& \ \Phi^\rho(\mu,t) \neq \Phi^\rho(\mu',t),$$
$\forall \mu \in \boldsymbol{B}^n, \exists \mu' \in \boldsymbol{B}^n, \forall \rho \in P_n, \exists t \in \boldsymbol{R}$,
$$\bigcup_{\xi \in (-\infty,t]} \rho(\xi) = (1,...,1) \ \& \ \Phi^\rho(\mu,t) \neq \Phi^\rho(\mu',t).$$

Remark 3.2 The prefixes 'p', 'n' are understood as 'possibly' and 'necessarily'. On the other hand the last two properties are the negations of the first two.

In [1] page 31 (R. Devaney is cited) the sensitive dependence on the initial conditions (SDIC) is defined like this. Let $(X,d)$ be a metric space. A dynamical system $(T,X,\Phi)$ has SDIC ($T = \boldsymbol{R}_+$ is the time set, $X$ is the state space and $\Phi: T \times X \to X$ is the flow) if $\varepsilon > 0$ exists such that for any $x \in X$ and any neighborhood $V$ of $x$, some $y \in V$ and some $t > 0$ exist with $d(\Phi_t(x), \Phi_t(y)) > \varepsilon$. In the rough topology of $X = \boldsymbol{B}^n$, the neighborhood $V$ of $\mu$ is $\boldsymbol{B}^n$ and $d(\Phi_t(x), \Phi_t(y)) > \varepsilon$ becomes $\Phi^\rho(\mu,t) \neq \Phi^\rho(\mu',t)$. We have interpreted 'some $t > 0$ exists' as $\exists t \in \boldsymbol{R}, \bigcup_{\xi \in (-\infty,t]} \rho(\xi) = (1,...,1)$.

*Theorem* 3.3 Let be $\Phi, \Upsilon: \boldsymbol{B}^n \to \boldsymbol{B}^n$ and we presume that $\Xi_\Phi$ and $\Xi_\Upsilon$ are conjugated. If $\Xi_\Phi$ is p-independent (n-independent) on the initial states then $\Xi_\Upsilon$ has the same property.

*Proof* If $\exists \mu \in \boldsymbol{B}^n, \forall \mu' \in \boldsymbol{B}^n, \exists \rho \in P_n, \forall t \in \boldsymbol{R}$,
$$\bigcup_{\xi \in (-\infty,t]} \rho(\xi) = (1,...,1) \Rightarrow \Phi^\rho(\mu,t) = \Phi^\rho(\mu',t)$$
and $H: \boldsymbol{B}^n \to \boldsymbol{B}^n, H' \in \Omega$ exist such that $\forall \nu \in \boldsymbol{B}^n$,
$$H \circ \Phi^\nu = \Upsilon^{H'(\nu)} \circ H,$$
i.e. $\forall \mu \in \boldsymbol{B}^n, \forall \rho \in P_n, \forall t \in \boldsymbol{R}$,

Șerban E. VLAD

$$H(\Phi^\rho(\mu,t)) = \Upsilon^{H'(\rho)}(H(\mu),t),$$

then

$$H'(\bigcup_{\xi\in(-\infty,t]}\rho(\xi)) = \bigcup_{\xi\in(-\infty,t]} H'(\rho(\xi))$$

$$= \bigcup_{\xi\in(-\infty,t]} H'(\rho)(\xi) = H'(1,...,1) = (1,...,1),$$

$$\Upsilon^{H'(\rho)}(H(\mu),t) = H(\Phi^\rho(\mu,t))$$

$$= H(\Phi^\rho(\mu',t)) = \Upsilon^{H'(\rho)}(H(\mu'),t)$$

hold.

*Theorem* 3.4 The following statements are equivalent:
a) the n-independence on the initial states;
b) $\Phi$ is the constant function.

*Proof* a)$\Rightarrow$ b) We suppose against all reason that $\Phi$ takes at least two different values, $\Phi(\mu) = \mu'' \neq \mu'' = \Phi(\mu')$. We choose in a) $\rho$ such that $\rho(t_0) = (1,...,1)$ and we have for $t = t_0$:

$$\Phi^\rho(\mu,t_0) = \Phi(\mu) = \mu'' \neq \mu'' = \Phi(\mu') = \Phi^\rho(\mu',t_0),$$

contradiction.

b)$\Rightarrow$ a) If $\Phi = \mu''$ is the constant function, we have that $\Phi^\rho(\mu,t)$ is of the form:

$$\Phi^\rho(\mu,t) = \mu\chi_{(-\infty,t_0)}(t) \oplus \Phi^{\alpha^0}(\mu)\chi_{[t_0,t_1)}(t) \oplus ...$$

$$... \oplus \Phi^{\alpha^0\cup...\cup\alpha^k}(\mu)\chi_{[t_k,t_{k+1})}(t) \oplus ...$$

with $\lim_{k\to\infty}\alpha^0\cup...\cup\alpha^k = (1,...,1)$. If $t\in[t_k,t_{k+1})$, $k\in N$ and $\alpha^0\cup...\cup\alpha^k = (1,...,1)$ we get

$$\Phi^{\alpha^0\cup...\cup\alpha^k}(\mu) = \Phi(\mu) = \mu'',$$

in other words a) holds under the form

$$\alpha^0\cup...\cup\alpha^k = (1,...,1) \Rightarrow \Phi^\rho(\mu,t) = \mu'' = \Phi^\rho(\mu',t)$$

*Example* 3.5 The system from Fig. 1 is p-independent on

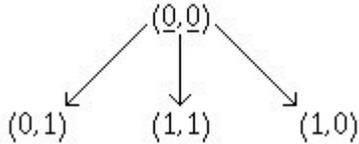

*Fig. 1* p-independence and p-dependence on the initial states.

the initial states. For this we can choose in Definition 3.1 $\mu = (0,0)$. This system is also p-dependent on the initial states, since $\Phi$ from that Figure is not constant.

*Example* 3.6 In Fig. 2 the system is n-independent on the initial states since $\Phi = (1,1)$ is the constant function.

*Example* 3.7 The system from Fig. 3 is n-dependent on the initial states. For this we can take in Definition 3.1 $\mu\in\{(0,0),(0,1)\}$, $\mu'\in\{(1,0),(1,1)\}$ and vice-versa.

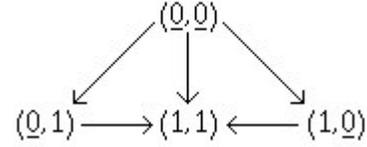

*Fig. 2* n-independence on the initial states.

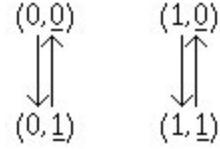

*Fig. 3* n-dependence on the initial states.

*Open problem* 3.8 Which is the relation between the p-independence on the initial states and the existence of the fixed points of $\Phi$?

*Definition* 3.9 The orbits of $\mu,\mu'\in B^n$ are called (temporally) p-separated if $\exists\rho\in P_n, \exists t\in R$,

$$\bigcup_{\xi\in(-\infty,t]}\rho(\xi) = (1,...,1) \,\&\, \Phi^\rho(\mu,t) \neq \Phi^\rho(\mu',t).$$

In this case we say that they are weakly p-separated if the set

$$\{t \mid \bigcup_{\xi\in(-\infty,t]}\rho(\xi) = (1,...,1) \,\&\, \Phi^\rho(\mu,t) \neq \Phi^\rho(\mu',t)\}$$

is bounded and else that they are strongly p-separated

$$\exists\rho\in P_n, \forall t\in R, \Phi^\rho(\mu,t) \neq \Phi^\rho(\mu',t).$$

The orbits of $\mu,\mu'\in B^n$ are atemporally p-separated if

$$\exists\rho\in P_n, Or_\rho(\mu)\cap Or_\rho(\mu') = \emptyset.$$

The n-separation of the orbits is defined by replacing $\exists\rho\in P_n$ with $\forall\rho\in P_n$.

*Remark* 3.10 The definition of this notion from [1] page 32 is the following one. Let $(X,d)$ be a metric space, $(T,X,\Phi)$ be a dynamical system and $\varepsilon > 0$. The orbits of the points $x,y\in X$ are $\varepsilon$-separated if $t\geq 0$ exists such that $d(\Phi_t(x),\Phi_t(y)) > \varepsilon$; they are $\varepsilon$-weakly separated if the set $\{t\geq 0 \mid d(\Phi_t(x),\Phi_t(y)) > \varepsilon\}$ is bounded, $\varepsilon$-recurrently separated if a sequence $(t_k)\subset T$ exists with $\lim_{k\to\infty} t_k = \infty$ such that $d(\Phi_{t_k}(x),\Phi_{t_k}(y)) > \varepsilon$ and $\varepsilon$-strongly separated if $t'\in T^+$ exists such that $d(\Phi_t(x),\Phi_t(y)) > \varepsilon$ for any $t\geq t'$. We see that in the definition of the $\varepsilon$-recurrently separated orbits we have in fact that $(t_k)\in Seq$ and this concept coincides in our case with the strong p-separated orbits.

Şerban E. VLAD

We conclude that $\Xi_\Phi$ is p-dependent (n-dependent) on the initial states if and only if $\forall \mu \in \boldsymbol{B}^n, \exists \mu' \in \boldsymbol{B}^n$ such that the orbits of $\mu, \mu'$ are p-separated (n-separated).

*Theorem* 3.11 We suppose that the systems $\Xi_\Phi, \Xi_\Upsilon$ are conjugated, with $\Phi, \Upsilon: \boldsymbol{B}^n \to \boldsymbol{B}^n$ and let $H, H'$ be like at Definition 2.14. If the orbits of $\mu, \mu' \in \boldsymbol{B}^n$ are for $\Xi_\Phi$ temporally p-separated (weakly p-separated, strongly p-separated), respectively atemporally p-separated, then the orbits of $H(\mu), H(\mu')$ are for $\Xi_\Upsilon$ temporally p-separated (weakly p-separated, strongly p-separated), respectively atemporally p-separated.

*Proof* We presume that $\exists \rho \in P_n$, $Or_\rho(\mu) \cap Or_\rho(\mu') = \emptyset$ and that $H: \boldsymbol{B}^n \to \boldsymbol{B}^n$, $H' \in \Omega$ exist such that $\forall \nu \in \boldsymbol{B}^n$,

$$H \circ \Phi^\nu = \Upsilon^{H'(\nu)} \circ H,$$

i.e. $\forall \mu \in \boldsymbol{B}^n, \forall \rho \in P_n, \forall t \in \boldsymbol{R}$,

$$H(\Phi^\rho(\mu, t)) = \Upsilon^{H'(\rho)}(H(\mu), t).$$

The fact that $\exists \rho \in P_n, \forall t \in \boldsymbol{R}, \forall t' \in \boldsymbol{R}$,

$$\Phi^\rho(\mu, t) \neq \Phi^\rho(\mu', t')$$

becomes $\exists \rho \in P_n, \forall t \in \boldsymbol{R}, \forall t' \in \boldsymbol{R}$,

$$H^{-1}(\Upsilon^{H'(\rho)}(H(\mu), t)) \neq H^{-1}(\Upsilon^{H'(\rho)}(H(\mu'), t'))$$

i.e.

$$\Upsilon^{H'(\rho)}(H(\mu), t) \neq \Upsilon^{H'(\rho)}(H(\mu'), t')$$

and finally

$$\exists \rho \in P_n, Or_{H'(\rho)}(H(\mu)) \cap Or_{H'(\rho)}(H(\mu')) = \emptyset$$

(for $\Xi_\Phi$).

*Example* 3.12 In Fig. 4 the existence of $\rho \in P_3$,

$$\rho(t) = (1,1,1)\chi_{\{t_0\}}(t) \oplus (0,0,1)\chi_{\{t_1\}}(t)$$
$$\oplus (1,1,0)\chi_{\{t_2\}}(t) \oplus (1,1,1)\chi_{\{t_3\}}(t)$$
$$\oplus (0,0,1)\chi_{\{t_4\}}(t) \oplus ...$$

shows that the orbits of $(0,1,1)$ and $(0,0,0)$ are weakly p-separated. At the same time the orbits $Or_\rho(1,1,0) = \{(1,1,0), (1,0,1), (1,0,0)\}$ and $Or_\rho(0,1,1) = \{(0,1,1)\}$ are atemporally p-separated.

*Example* 3.13 The orbits of $(0,0)$ and $(1,0)$ are atemporally n-separated in Fig. 3.

*Definition* 3.14 A point $\mu \in \boldsymbol{B}^n$ is called weakly p-transitive, respectively strongly p-transitive, respectively n-transitive if

$$\forall \mu' \in \boldsymbol{B}^n, \exists \rho \in P_n, \exists t \in \boldsymbol{R}, \Phi^\rho(\mu, t) = \mu',$$
$$\exists \rho \in P_n, \forall \mu' \in \boldsymbol{B}^n, \exists t \in \boldsymbol{R}, \Phi^\rho(\mu, t) = \mu',$$

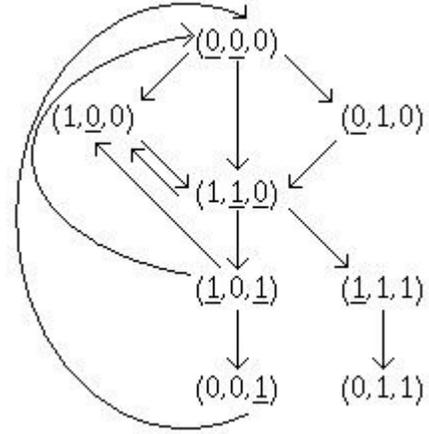

Fig. 4 The separation of the orbits.

$$\forall \mu' \in \boldsymbol{B}^n, \forall \rho \in P_n, \exists t \in \boldsymbol{R}, \Phi^\rho(\mu, t) = \mu'.$$

*Remark* 3.15 The transitive points are defined in [1], page 22 by the fact that the orbit of such a point is dense in the state space. The transitivity of $\mu$ states that all the points of $\boldsymbol{B}^n$ are possibly or necessarily accessible from $\mu$.

*Example* 3.16 The point $(1,1)$ in Fig. 5 is weakly p-transitive.

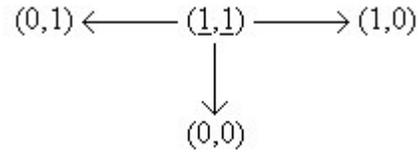

Fig. 5 The point $(1,1)$ has the property of weak p-transitivity.

*Example* 3.17 In Fig. 6 the point $(1,0)$ fulfills the pro perty of strong p-transitivity by taking $\rho \in P_2$,

$$\rho(t) = (0,1)\chi_{\{t_0\}}(t) \oplus (1,0)\chi_{\{t_1\}}(t)$$
$$\oplus (0,1)\chi_{\{t_0\}}(t) \oplus (1,0)\chi_{\{t_1\}}(t) \oplus ...$$

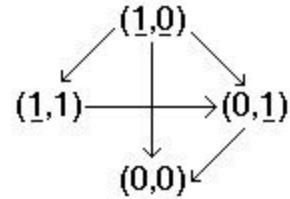

Fig. 6 The point $(1,0)$ has the property of strong p-transitivity.

*Example* 3.18 The point $(1,0)$ has in Fig. 7 the property of n-transitivity.

Șerban E. VLAD

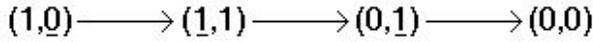

Fig. 7 The point $(1,0)$ has the property of n-transitivity.

*Definition* 3.19 The set $A \in P^*(B^n)$ is weakly p-transitive, respectively strongly p-transitive, respectively n-transitive if

$$\forall \mu \in A, \exists \rho \in P_n, Or_\rho(\mu) = A,$$
$$\exists \rho \in P_n, \forall \mu \in A, Or_\rho(\mu) = A,$$
$$\forall \mu \in A, \forall \rho \in P_n, Or_\rho(\mu) = A.$$

If previously $A = B^n$, then the system $\Xi_\Phi$ (or $\Phi$) is called weakly p-transitive, respectively strongly p-transitive, respectively n-transitive.

*Remark* 3.20 In [2], page 6 the set $A \subset X$ is called topologically transitive for $(T, X, \Phi)$ (Wiggins and Georgescu are cited) if it is invariant, closed and for any open sets $U, V \subset A$, $\exists t \in T$ such that $\Phi_t(U) \cap V \neq \emptyset$.

The open sets have been interpreted as points $\mu, \mu' \in A$ in the discrete topology of $X = B^n$. The transitivity of a set $A$, any of the three possibilities from Definition 3.19, consists for $\mu \in A$ and $\rho \in P_n$ in two requests: the request of invariance $Or_\rho(\mu) \subset A$ and the request of accessibility $A \subset Or_\rho(\mu)$.

In [1], page 22 (E. Petrisor is cited) and [2], page 3 it is stated that the topological transitivity of a discrete time dynamical system $(N, X, \Phi)$ where $\Phi : X \to X$ is continuous consists in the existence of $x_0 \in X$ such that $\overline{Or(x_0)} = X$. In [1], page 23 the minimality of $\Phi$ is defined (H. Furstenberg and R. N. Gologan are cited) by the fact that $\forall x_0 \in X$ fulfills $\overline{Or(x_0)} = X$. We preferred to identify (unlike [1], [2] where the transitivity makes use of $\exists x_0 \in X$) the transitivity of a system with its minimality.

*Example* 3.21 In Fig. 8 the set $A = \{(0,0),(1,0)\}$ is weakly p-transitive.

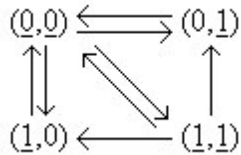

Fig. 8 The set $A = \{(0,0),(1,0)\}$ is weakly p-transitive.

*Example* 3.22 In Fig. 9 the set $A = \{(0,0),(1,1)\}$ is strongly p-transitive, with $\rho \in P_2$,

$$\rho(t) = (1,1)\chi_{\{t_0\}}(t) \oplus (1,1)\chi_{\{t_1\}}(t) \oplus (1,1)\chi_{\{t_2\}}(t) \oplus ...$$

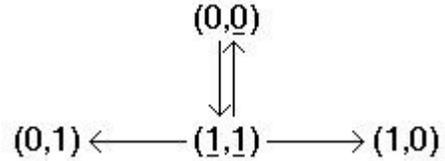

Fig. 9 The set $A = \{(0,0),(1,1)\}$ is strongly p-transitive.

*Example* 3.23 In Fig. 3 the sets $A = \{(0,0),(0,1)\}$ and $A = \{(1,0),(1,1)\}$ are n-transitive.

*Example* 3.24 The system from Fig. 10 is weakly p-transitive.

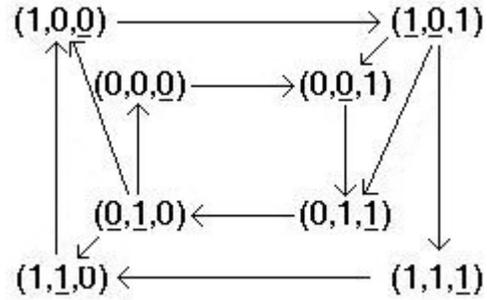

Fig. 10 Weakly p-transitive system.

Example 3.25 The property of strong p-transitivity is fulfilled by the system from Fig. 11, with $\rho \in P_2$,

$$\rho(t) = (0,1)\chi_{\{t_0\}}(t) \oplus (1,0)\chi_{\{t_1\}}(t)$$
$$\oplus (0,1)\chi_{\{t_0\}}(t) \oplus (1,0)\chi_{\{t_1\}}(t) \oplus ...$$

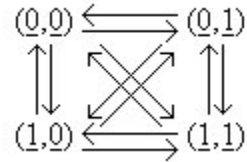

Fig. 11 Strongly p-transitive system.

*Example* 3.26 The system from Fig. 12 is n-transitive.

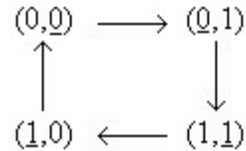

Fig. 12 n-transitive system.

*Theorem* 3.27 Let be $\Phi, \Upsilon : B^n \to B^n$ and we presume that $\Xi_\Phi$ and $\Xi_\Upsilon$ are conjugated. If $\Xi_\Phi$ is weakly p-transitive (strongly p-transitive, n-transitive), then $\Xi_\Upsilon$ is weakly p-transitive (strongly p-transitive, n-transitive).

Şerban E. VLAD

*Proof* The hypothesis states the existence of the bijections $H : \boldsymbol{B}^n \to \boldsymbol{B}^n$ and $H' \in \Omega$ such that $\forall \nu \in \boldsymbol{B}^n$,
$$H \circ \Phi^\nu = \Upsilon^{H'(\nu)} \circ H,$$
i.e. $\forall \mu \in \boldsymbol{B}^n$, $\forall \rho \in P_n$, $\forall t \in \boldsymbol{R}$,
$$H(\Phi^\rho(\mu,t)) = \Upsilon^{H'(\rho)}(H(\mu),t).$$
The request of weak p-transitivity $\forall \mu \in \boldsymbol{B}^n$, $\forall \mu' \in \boldsymbol{B}^n$, $\exists \rho \in P_n$, $\exists t \in \boldsymbol{R}$,
$$\Phi^\rho(\mu,t) = \mu'$$
may be rewritten under the form $\forall \mu \in \boldsymbol{B}^n$, $\forall \mu' \in \boldsymbol{B}^n$, $\exists \rho \in P_n$, $\exists t \in \boldsymbol{R}$,
$$\Phi^{H'^{-1}(\rho)}(H^{-1}(\mu),t) = H^{-1}(\mu')$$
wherefrom we get $\forall \mu \in \boldsymbol{B}^n$, $\forall \mu' \in \boldsymbol{B}^n$, $\exists \rho \in P_n$, $\exists t \in \boldsymbol{R}$,
$$\Upsilon^\rho(\mu,t) = \Upsilon^{H'(H'^{-1}(\rho))}(H(H^{-1}(\mu)),t)$$
$$= H(\Phi^{H'^{-1}(\rho)}(H^{-1}(\mu),t)) = H(H^{-1}(\mu')) = \mu'.$$
The other statements are proved similarly.

*Theorem* 3.28 If the system $\Xi_\Phi$ is weakly p-transitive (n-transitive), then it p-depends (n-depends) on the initial states.

*Proof* We prove the first statement of the Theorem and we suppose against all reason that $\Xi_\Phi$ is not p-dependent on the initial states and then it is n-independent on the initial states meaning that $\Phi$ is the constant function $\Phi = \mu$, see Theorem 3.4. In this situation we have
$$\forall \rho \in P_n, Or_\rho(\mu) = \{\mu\}$$
representing a contradiction with the hypothesis of p-transitivity stating that
$$\exists \rho \in P_n, Or_\rho(\mu) = \boldsymbol{B}^n.$$
The second statement of the Theorem was not proved so far.